\documentclass[fleqn,12pt]{wlscirep}
\usepackage[utf8]{inputenc}
\usepackage[T1]{fontenc}
\title{Corner Topology Makes Woven Baskets into Stiff, yet Resilient Metamaterials}

\usepackage{caption}
\usepackage{helvet}  % For Arial/Helvetica font
\makeatletter
\renewcommand{\@makecaption}[2]{%
  {\fontfamily{phv}\selectfont\footnotesize\textbf{#1.} #2\par}%
}
\makeatother % added by Wayne, to make the caption text font style 'Helvetica' and the font size 'footnotesize'. 

\author[1]{Guowei Wayne Tu}
\author[1,2,*]{Evgueni T. Filipov}
\affil[1]{Deployable and Reconfigurable Structures Laboratory, Department of Civil and Environmental Engineering, University of Michigan, Ann Arbor, MI, USA}
\affil[2]{Deployable and Reconfigurable Structures Laboratory, Department of Mechanical Engineering, University of Michigan, Ann Arbor, MI, USA}
\affil[*]{E-mail: filipov@umich.edu}


\begin{abstract}
Basket weaving is a traditional craft used to create practical three-dimensional (3D) structures. While the geometry and aesthetics of baskets have received considerable attention, the underlying mechanics and modern engineering potential remain underexplored. This work shows that 3D woven structures offer similar stiffness yet substantially higher resilience than their non-woven continuous counterparts. We explore corner topologies that serve as building blocks to convert 2D woven sheets into 3D metamaterials that can carry compressive loads. Under small deformations, the woven corners exhibit axial stiffness similar to continuous structures because the woven ribbons are engaged with in-plane loads. Under large deformations, the woven corners can be compressed repeatedly without plastic damage because ribbons can undergo elastic local buckling. We present a modular platform to assemble woven corners into complex spatial metamaterials and demonstrate applications including damage-resilient robotic systems and metasurfaces with tailorable deformation modes. Our results explain the historic appeal of basket weaving, where readily available ribbons are crafted into 3D structures with comparable stiffness yet far superior resilience to continuous systems. The modular assembly of woven metamaterials can further revolutionize design of next-generation automotive components, consumer devices, soft robots, and more where both resilience and stiffness are essential. 
\end{abstract}
\begin{document}

\flushbottom
\maketitle
\thispagestyle{empty}

\section*{Introduction}
Weaving is the craft of converting one-dimensional (1D) fibers or ribbons into two-dimensional (2D) composite materials through intentional physical entanglement \cite{lan2023woven}. This entanglement offers mechanical redundancy where even if constituent fibers are damaged, the entire hierarchical system can still function properly \cite{wang2017smart,parsons2010impact,parsons2013modeling}. Among various weaving patterns, plain weaving is the simplest and most common where perpendicularly oriented strands are interlaced in a one-over-one-under configuration \cite{august2020self,must2019variable,maziz2017knitting,zhang2017rugged,zhao2021soft}. Plain-woven composites can offer near-continuous material coverage and a potentially impermeable surface \cite{backer1951relationship}. Such 2D textiles have been used since antiquity to create clothes, shoes, and furniture \cite{granger1982weaving}, while modern-day innovation has focused on creating smart and electronic textiles for biomechanical sensing \cite{wu2019silk,zhang2022elastic,shi2021large}, energy harvesting \cite{huang2020fiber,chai2016tailorable,chen2016micro}, and soft robotic actuation \cite{xiong2021functional,buckner2024untethered,buckner2020roboticizing}. 

While 2D woven systems offer many advantages, they cannot function as structures because the thin planar surfaces buckle under small compressive loads \cite{hu2019buckled,suzuki2016wrinkles,rajesh2017experimental}. To address this issue, the craft of basketry, recorded since at least 7500 BCE \cite{martinez2023earliest}, transforms the 2D woven sheet into a self-reinforcing three-dimensional (3D) structure that retains its shape and can carry compressive loads. For the classical plain weave, this transition to 3D space can be achieved by connecting three perpendicular woven patches into a conically curved triangle with the same topology as a cube corner (see Fig. \ref{fig:Intro}a). Related corner topologies can be achieved with tri-axial sparse weaving, and prior work has used this principle to design complex 3D woven geometries such as domes \cite{poincloux2023indentation}, spheres \cite{song2023buckling}, and tori \cite{baek2021smooth,ren20213d}. Although tri-axial sparse weaving enables the realization of such sophisticated 3D geometries, it requires more advanced fabrication techniques than plain weaving \cite{lewandowska2017triaxial,peng2024switchable}, and the sparsity in the weave pattern results in a flexible discontinuous surface with large gaps between ribbons. 

In this work, we explore the simply fabricated, continuous corner topologies created using bi-axial plain, dense weaving. We generalize the topology of the natural basket corner in Fig. \ref{fig:Intro}a to obtain a woven corner family where each variation has a different topological number (Fig. \ref{fig:Intro}b and c). We then show how these woven corners can be assembled as modular building blocks to create 3D structures with various spatial geometries (Fig. \ref{fig:Intro}d). Beyond their geometric versatility, these woven corners provide surprisingly high stiffness and resilience to the final structural system. Using theory, numerical simulations, and experiments, we demonstrate that modular woven structures can be nearly as stiff as their continuous counterparts, yet do not suffer plastic damage when global buckling occurs (Fig. \ref{fig:Intro}e, see Supplementary Video S1). These findings help explain why our ancestors split planar sheets into fibers and wove them into baskets rather than directly using the continuous sheets in their design \cite{martinez2023earliest}. Our approach enables the fabrication of 3D structures from 1D slender elements without external supports, and more importantly, the weaving significantly enhances durability without sacrificing stiffness. Our findings further inspire the design and fabrication of 3D functional metastructures and metamaterials that are lightweight, yet stiff for load bearing and resilient to damage. We give examples including woven robotic systems that are resilient to accidental overload and woven metasurfaces that can be tailored to be stiff or flexible.

\section*{Woven Corner Units and Modular Assembly}
\begin{figure}[!htb] % Use this if the figure floating is not ideal
% \begin{figure*}[!htbp] 
\centering
\includegraphics[width=17.8cm]{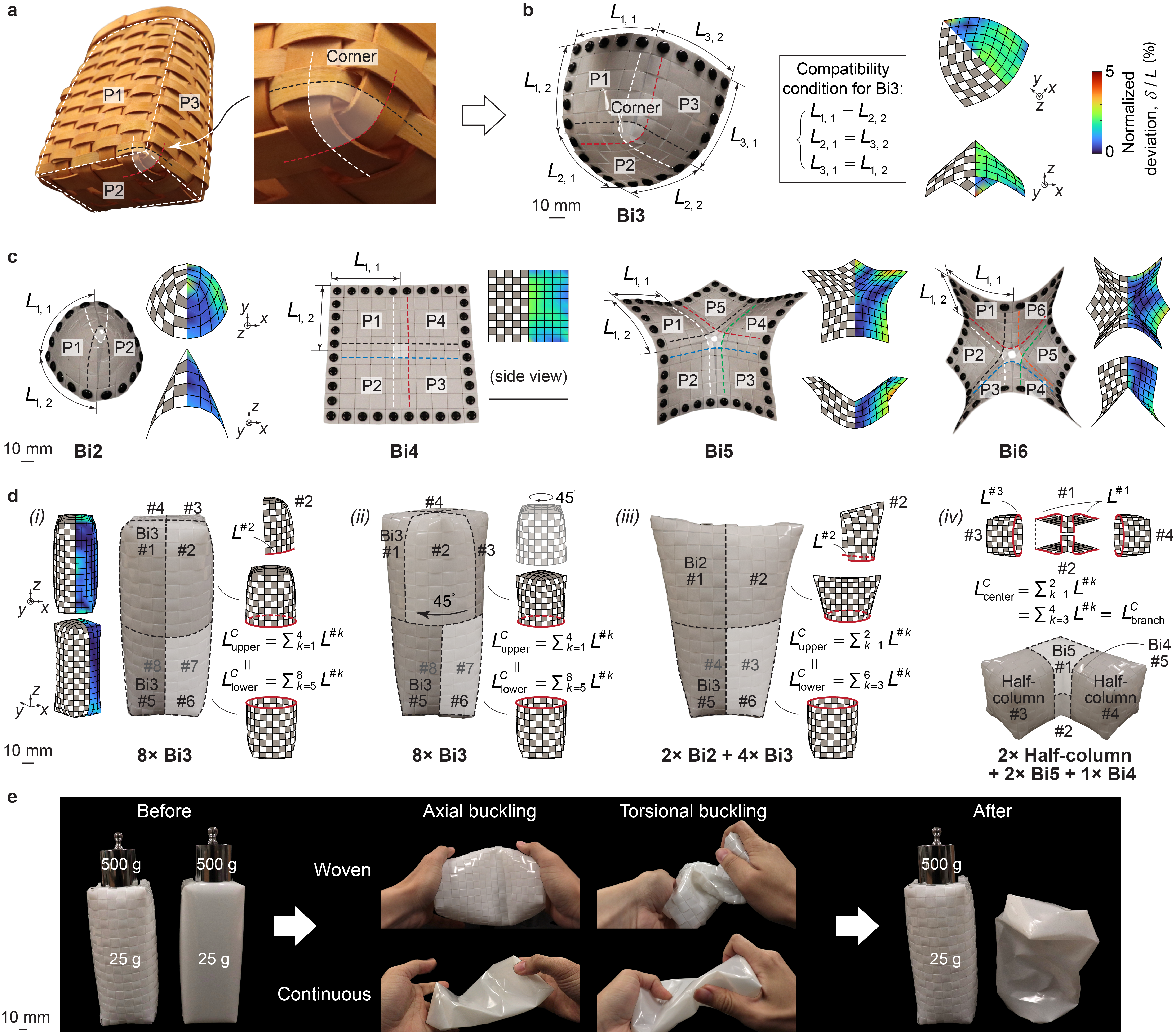}
\caption{\textbf{Woven corner units and assemblies}. \textbf{a}, A Bi3 corner unit connects three rectangular woven patches to form a basket. \textbf{b}, Physical prototype of the Bi3 corner with its simulated shape on the right. The colormap represents the normalized deviation between the simulated and scanned geometries $\delta/\overline{L}$ where $\delta$ is the deviation (Hausdorff distances at each node \cite{huttenlocher1993comparing}) and $\overline{L}$ is the average of all patch lengths. \textbf{c}, Physical prototypes of Bi2, Bi4, Bi5, and Bi6 corners with the corresponding simulated shapes. The dashed colored lines drawn on the photos of prototypes in \textbf{b} and \textbf{c} indicate the paths of the ribbons that meet at the centers of each corner. \textbf{d}, Physical prototypes and simulated shapes of (\textit{i}) A woven column made by interweaving eight identical Bi3 corners; (\textit{ii}) A variant of the woven column in (\textit{i}) with the upper portion rotated by 45$^{\circ}$; (\textit{iii}) A variant of the woven column in (\textit{i}) with the upper portion replaced by two Bi2 corners; (\textit{iv}) An elbow link where two half-column modules are connected using two Bi5 corners and one Bi4 unit. \textbf{e}, A physical comparison where a load-bearing woven column fully recovers after axial and torsional buckling, while a geometrically equivalent continuous column experiences plastic damage and cannot recover (see Supplementary Video S1).}\label{fig:Intro}
\end{figure}

We interweave three groups of bi-axially arranged Mylar\textsuperscript{\textregistered} polyester ribbons of the same dimensions (135$\times$13$\times$0.1905 mm) in three-dimensional space to obtain a \textit{Bi3 corner} that has the same topology as the basket corner (Fig. \ref{fig:Intro}b). Each group has five parallel ribbons, and split pins are placed at the external crossings to secure the boundary (see details on fabrication in Supplementary Fig. S1 and Section S1A). Here, `Bi' means bi-axial, and `3' means that the corner has three patches (P1, P2, P3 in Fig. \ref{fig:Intro}b). For rectangular woven patches to be compatible, we need to make the sides of adjacent rectangles pairwise equal in length (for example, $L_{1, 1}=L_{2, 2}$ where $L_{i,j}$ is the length of panel $i$ on side $j$). This \textit{compatibility condition} is given as a system of equations in Fig. \ref{fig:Intro}b. Applying similar compatibility conditions to two, four, five, and six rectangular patches, we obtain generalized woven corners including Bi2, Bi4, Bi5, and Bi6 (Fig. \ref{fig:Intro}c). The Bi3 corner connects three patches and results in a partial conical shape where the sector angle is 270$^{\circ}$, while the Bi2 corner connects only two patches and results in a sharper partial cone where the sector angle is 180$^{\circ}$. The Bi4 is a trivial shape, which is a flat sheet where the sector angle is 360$^{\circ}$. Combining more patches results in saddle-like surfaces such as the Bi5 and Bi6 corners where the sector angles are 450$^{\circ}$ and 540$^{\circ}$, respectively. We simulate the rest shapes of these corners using a self-morphing solver based on a bar and hinge model \cite{tu2024origami,filipov2017bar,zhu2021rapid}, and the simulated shapes have excellent agreement with 3D scans of the physical prototypes (Fig. \ref{fig:Intro}b and c, see Supplementary Sections S2 and S3 for details on simulation and scanning). 

Using similar length compatibility conditions between adjacent patches, we can use the woven corners as building blocks to design more complex woven topologies in a modular manner. Figure \ref{fig:Intro}d(\textit{i}) shows a column-like structure made by interweaving eight identical Bi3 corners. Here, four Bi3 corners \#1--4 form the upper part and the other four \#5--8 form the lower part. The compatibility condition can then be simplified based on the entire column, where the combined length of all units in the upper part is equal to the combined length of the lower part, written as $L_{\mathrm{upper}}^{C} = \sum\nolimits_{k = 1}^4 {L^{\# k}}=L_{\mathrm{lower}}^{C} = \sum\nolimits_{k = 5}^8 {L^{\# k}}$ where $L^{\#k}$ are now only the relevant lengths of each corner building block. This more general system-level compatibility condition still holds when we rotate the upper part of the column by a specific angle or replace the upper part with a combination of two Bi2 units, as shown in Fig. \ref{fig:Intro}d(\textit{ii}) and (\textit{iii}) respectively. We can also connect two half-column modules with two Bi5 units in the center to form an elbow link structure (Fig. \ref{fig:Intro}d(\textit{iv})). Using the basic corner units, we can use a growth-inspired strategy to design and fabricate arbitrary complexity, where the only rule for connecting any two units is the `local edge length compatibility condition'. Figure \ref{fig:Intro}d shows structures with closed surfaces, and we give more examples of woven assemblies with open surfaces in Supplementary Fig. S4 and Section S4, where corners of higher orders such as Bi7 and Bi8 are also used. The resulting woven structures are nearly as stiff as their continuous counterparts, yet woven structures are different in that they do not suffer plastic damage under extreme load (Fig. \ref{fig:Intro}e and see Supplementary Video S1). 

\section*{Resilience Enabled by Local Buckling of Ribbons}
\begin{figure}
% \begin{figure}[!htb] % Use this if the figure floating is not ideal
\centering
\includegraphics[width=17.8cm]{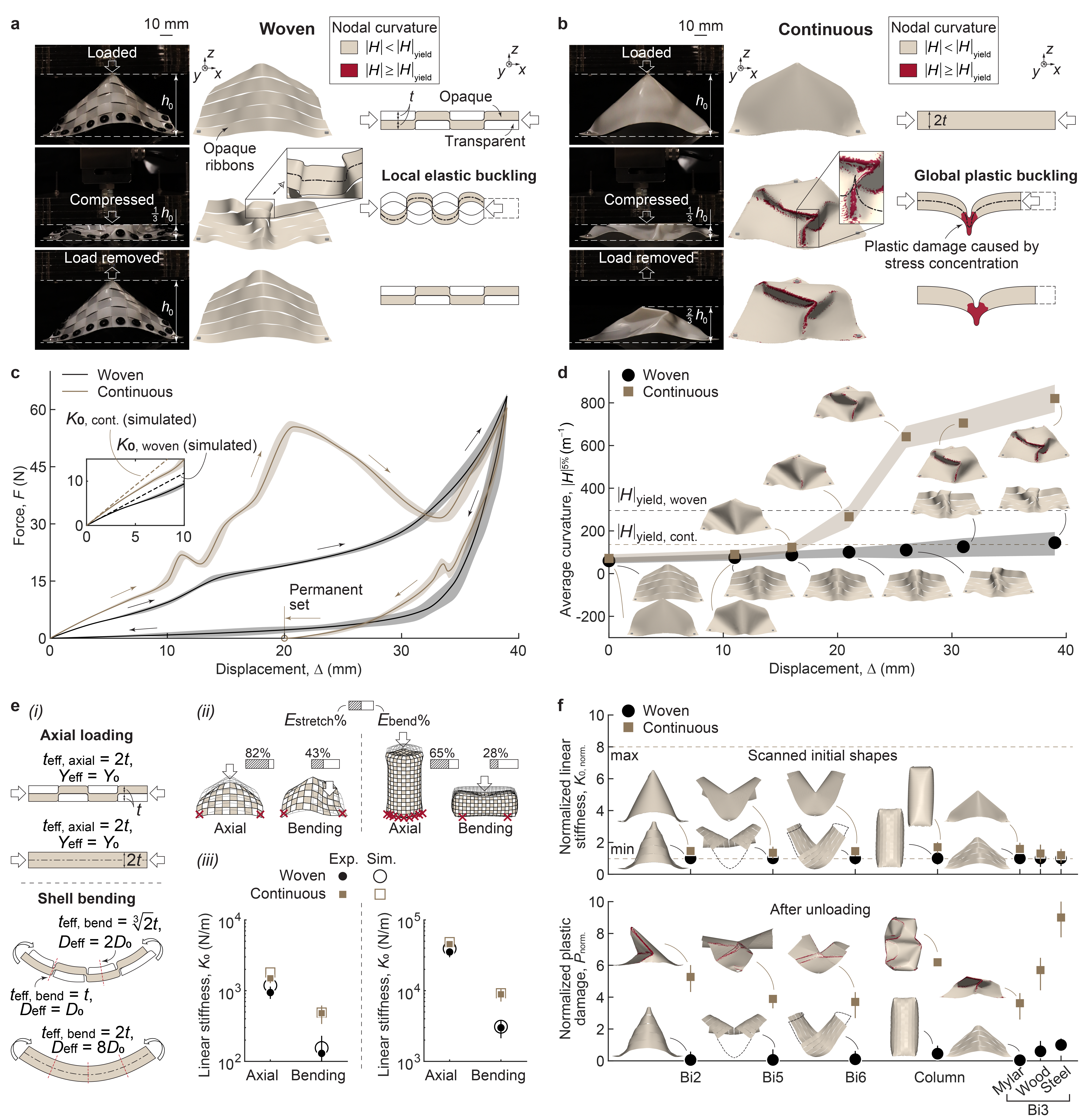}
\caption{\textbf{Stiffness and resilience of woven corners.} \textbf{a} and \textbf{b} show photographs (left panels), scanned shapes (middle panels), and sketched mechanical behavior (right panels) of the woven and continuous Bi3 corners when we initially load the corners, compress them, and remove the load (see Supplementary Video S2). \textbf{c}, Force--displacement curves of the woven and continuous Bi3 corners. The inset shows a comparison between the experimental and simulated linear stiffness $K_0$. \textbf{d}, Average of the top 5\% nodal curvature values $|H|^{\overline{\mathrm{5\%}}}$ of the woven and continuous Bi3 corner surfaces during the vertical loading process with snapshots of scanned deformed shapes. \textbf{e}, (\textit{i}) The effective thicknesses and effective moduli of two woven ribbons and a continuous shell are the same in axial loading (${t_{{\rm{eff,  \ axial}}}}$, $Y_{\mathrm{eff}}$), however they are lower for the woven system in lateral bending ($t_{\mathrm{eff}, \ \mathrm{bend}}$, $D_{\mathrm{eff}}$); (\textit{ii}) Simulated linearly deformed woven corner and woven column under axial loading and lateral bending with the corresponding stretching--bending energy ratios. The checkerboard-colored surfaces are the deformed configurations where the deformation is scaled up ($\times $200); the gray frames are initial undeformed shapes; the red crosses indicate the boundary nodes where the $x$, $y$, $z$ coordinates are fixed; a unit load is used in all simulations; (\textit{iii}) Linear stiffness of the woven corner and woven column under axial loading and lateral bending. \textbf{f}, Normalized linear stiffness and normalized plastic damage of woven and continuous Bi2, Bi3, Bi5, Bi6 corners and the column assemblies. The Bi3 corners are fabricated using Mylar\textsuperscript{\textregistered}, wood, and steel, while the others are only fabricated using Mylar\textsuperscript{\textregistered}. The solid lines in \textbf{c} and solid dots in \textbf{d}, \textbf{e}, \textbf{f} represent the average results of tests of five samples fabricated in the same manner, while the shaded areas in \textbf{c} and \textbf{d} and the vertical lines on dots in \textbf{e} and \textbf{f} represent the error ranges where the upper boundaries are the highest values and the lower boundaries are the lowest values in the five tests.}\label{fig:Mechanics}
\end{figure}

We first explore the resilience of woven structures by conducting load tests on Bi3 corners which are the most fundamental modules in woven assemblies. Figure \ref{fig:Mechanics}a shows the test setup where a Bi3 corner is clamped at its three vertices and is subject to vertical compression between two flat plates. For comparison, we conduct the same test on a corner that has the same geometry but is made of a continuous polyester sheet (Fig. \ref{fig:Mechanics}b, see Supplementary Section S1B for fabrication). The thickness of the continuous corner (0.3556 mm, $\approx2t$) is twice the thickness of the individual ribbons (0.1905 mm, $t$) in the woven corner so that the same amount of material is used for both corners. 

When the displacement is small ($\Delta \le$ 10 mm), both corners exhibit linear force--displacement behavior (Fig. \ref{fig:Mechanics}c). As the displacement increases, the continuous corner experiences global buckling followed by a gradual force decrease at $\Delta=$ 20 mm, while the woven corner has lower forces that monotonically increase. Both corners have similar force--displacement behaviors when they are compressed into the densification region where $\Delta\ge$ 30 mm (Fig. \ref{fig:Mechanics}c). As we remove the load, the woven corner recovers to its initial shape with the force dropping up until a zero displacement $\Delta=0$ mm (Fig. \ref{fig:Mechanics}c and Fig. \ref{fig:Mechanics}a, left panel, and see Supplementary Video S2). On the other hand, the continuous corner does not recover fully and experiences a permanent set of $\Delta =$ 20 mm (Fig. \ref{fig:Mechanics}c and Fig. \ref{fig:Mechanics}b, left panel). 

We scan the deformed structures at multiple points during the test to explain why these two corners are affected so differently by extreme loading. For woven corners, we interweave transparent and opaque ribbons so that we can explicitly scan the deformed geometry of the opaque ribbons that run in one direction without the confounding effect of the ribbons that run in the perpendicular direction (Fig. \ref{fig:Mechanics}a). Using the scanned surfaces, we calculate the absolute nodal curvature $|H|$ over the woven ribbons and the continuous corners, and we obtain the corresponding strains (see Supplementary Section S3 for details). We color the areas red where $|H|$ is higher than the yield curvature $|H|_{\mathrm{yield}}$ for all scanned surfaces in Fig. \ref{fig:Mechanics}, and $|H|_{\mathrm{yield}}$ is a threshold calculated based on the yield stress of the material (see Supplementary Section S3D for details). To further evaluate the strains within the `dangerous region' where plastic damage is most likely to occur, we compute the average curvature $|H|^{\overline{\mathrm{5\%}}}$ which is the average of the top 5\% of nodal curvature values over the entire surface. 

Under large deformations, the continuous corner exhibits global buckling with stress concentrations that exceed the yield stress ($\Delta>$ 20 mm, Fig. \ref{fig:Mechanics}d). The resulting plastic deformation is why the continuous corner is unable to recover. In contrast, the ribbons in the woven corner experience a segmented local elastic buckling that deconcentrates stresses (Fig. \ref{fig:Mechanics}a). The curvature of the ribbons increases steadily but never exceeds the yield curvature (Fig. \ref{fig:Mechanics}d), which is why the woven corner recovers to its initial shape. If the woven corner is re-loaded, it exhibits a nearly identical behavior and is able to dissipate as much energy as during its first load cycle (which is the area enclosed by the force--displacement curve, see Supplementary Fig. S5 and Video S2). Thus, while the continuous system dissipates more energy in one load cycle, the woven system can be a reusable energy absorber that undergoes the same absorption sequence multiple times \cite{fu2019programmable,chen2021reusable,jiang2023suture}. The yield curvature of the woven corner $|H|_{\mathrm{yield, \ woven}}$ is twice that of the continuous corner $|H|_{\mathrm{yield, \ cont.}}$ because the thickness of the ribbons ($t$) is half the thickness of the continuous shell ($2t$) (Fig. \ref{fig:Mechanics}d), but even if we double the thickness of the ribbons, local buckling still protects the woven corner from plastic damage (Supplementary Fig. S6 and Section S5B).

% \begin{figure*}

\newgeometry{left=1.7cm,right=1.7cm,top=1.3cm,bottom=2cm}
% added by Wayne (to modify the margin size for a specific page/figure)
\clearpage
\restoregeometry % after modification of the margin size, remember to change it back!! (added by Wayne)

\section*{Stiffness Achieved Through Stretching of Ribbons}
We define the experimental linear stiffness $K_0$ as the slope of a straight line that fits the force--displacement curve within the linear range ($\Delta \le$ 10 mm, Fig. \ref{fig:Mechanics}c). We also simulate $K_0$ using a linear analysis of the rest shape in Fig. \ref{fig:Intro}b, and we obtain a comparable result when the displacement is small. As the displacement increases, we observe a slight deviation of the simulated stiffness values from the experimental ones because our analysis is based on the assumption of infinitesimal deformations (the inset of Fig. \ref{fig:Mechanics}c, and see Supplementary Section S2B for details on simulation).

In axial compression, the woven corner gives us a stiffness close to that of the continuous corner ($K_{0, \ \mathrm{cont.}}/K_{0, \ \mathrm{woven}} \approx 1.5$, see the inset of Fig. \ref{fig:Mechanics}c). Here, we say that the stiffness is close as long as $K_{0, \ \mathrm{cont.}}/K_{0, \ \mathrm{woven}} < 2$, in contrast to the worst-case theoretical limit of 8, which will be discussed later. The bulk stiffnesses $K_0$ are close because the woven ribbons and the continuous shell have the same effective stretching modulus $Y_{\mathrm{eff}}$. According to the calculated stretching--bending energy ratio of the deformed configuration, the woven ribbons are mainly engaged in axial stretching and compression instead of lateral bending when the woven corner is loaded axially, ($E_{\mathrm{stretch}}=82\%$, Fig. \ref{fig:Mechanics}e(\textit{ii}), left panel, and see Supplementary Section S2A for details on energy calculation). In this case, the bulk stiffness of the woven system $K_0$ is mainly affected by the stretching stiffness of the material $K_A$, and $K_A$ scales linearly with the effective stretching modulus $Y_{\mathrm{eff}}$ (see Supplementary Eq. S2 and Section S2 for details). Therefore, the bulk stiffness $K_0$ scales approximately linearly with material modulus $Y_{\mathrm{eff}}$ ($K_0 \propto Y_{\mathrm{eff}}$), which is why an equal $Y_{\mathrm{eff}}$ leads to a close enough $K_0$ in axial compression. Here, we compute the effective stretching modulus using $Y_{\mathrm{eff}}=Et_{\mathrm{eff}, \ \mathrm{axial}}$ where $E$ is the Young's modulus and $t_{\mathrm{eff}, \ \mathrm{axial}}$ is the effective axial thickness. As the top panel of Fig. \ref{fig:Mechanics}e(\textit{i}) shows, without buckling or out-of-plane deformation, the woven ribbons and the continuous shell have the same effective axial thickness $t_{\mathrm{eff}, \ \mathrm{axial}}=2t$ and thus the same effective stretching modulus $Y_{\mathrm{eff}}$. 

On the other hand, in lateral bending where we load the woven corner in the middle of a patch, we pay a larger penalty and obtain a lower stiffness because of weaving ($K_{0, \ \mathrm{cont.}}/K_{0, \ \mathrm{woven}} \approx 3.5$, Fig. \ref{fig:Mechanics}e(\textit{ii}) and (\textit{iii}), left panels). This reduction occurs because the effective bending modulus of the woven ribbons is only $1/4$ to $1/8$ that of the continuous shell. Different from the axial loading case, in lateral bending, the woven ribbons are mainly engaged in bending instead of stretching ($E_{\mathrm{stretch}}=43\%$, Fig. \ref{fig:Mechanics}e(\textit{ii}), left panel). In this case, the bulk stiffness of the woven system $K_0$ is mainly affected by the bending stiffness of the material $K_B$, and $K_B$ scales linearly with the effective bending modulus $D_{\mathrm{eff}}$ (see Supplementary Eq. S1 and Section S2 for details). Therefore, the bulk stiffness $K_0$ scales almost linearly with material modulus $D_{\mathrm{eff}}$ ($K_0 \propto D_{\mathrm{eff}}$), which is why the lower $D_{\mathrm{eff}}$ of the woven system leads to a much lower $K_0$ in lateral bending. Here, we compute the effective bending modulus using $D_{\mathrm{eff}}=Et_{\mathrm{eff, \ bend}}^{3}/[12(1-\nu^2)]$ where $E$ is the Young's modulus, $\nu$ is the Poisson's ratio, and $t_{\mathrm{eff}, \ \mathrm{bend}}$ is the effective bending thickness. As the bottom panel of Fig. \ref{fig:Mechanics}e(\textit{i}) shows, along the ribbon spacing where only one ribbon gets bent, we have $t_{\mathrm{eff}, \ \mathrm{bend}}=t$, which makes the effective bending modulus $1/8 $ that of the continuous shell where $t_{\mathrm{eff}, \ \mathrm{bend}}=2t$; while in other regions where two perpendicular ribbons overlap, we have $t_{\mathrm{eff}, \ \mathrm{bend}}=\sqrt[3]{2}t$, which makes the effective bending modulus $1/4$ that of the continuous shell---the $t_{\mathrm{eff}, \ \mathrm{bend}}$ in this case is $\sqrt[3]{2}t$ instead of $2t$ because two non-rigidly bonded beams only result in a linear superposition of individual bending moduli of each beam \cite{Hibbeler2016Mechanics,peng2024analytic}. 

We observe the same phenomena where the axial stiffness is similar and the bending stiffness is penalized when we conduct axial compression and three-point bending of the woven column assembly (Fig. \ref{fig:Mechanics}e(\textit{ii}) and (\textit{iii}), right panels). Note that while weaving does result in a decrease in stiffness compared to continuous systems, the decrease is minimal if the structure is loaded axially, and even in bending the decrease is still less than the worst-case theoretical estimate. For example, in bending of the Bi3 corner, $K_{0, \ \mathrm{cont.}}/K_{0, \ \mathrm{woven}} \approx 3.5$ while the worst-case theoretical estimate is $K_{0, \ \mathrm{cont.}}/K_{0, \ \mathrm{woven}} = 8$. 

The high stiffness and enhanced resilience are two general properties that exist for the entire family of woven corner units (Bi2--Bi6) and the woven column assembly. Figure \ref{fig:Mechanics}f shows the results where a normalized linear stiffness $K_{\mathrm{0, \ norm.}}$ is defined as $K_{\mathrm{0}} / K_{\mathrm{0, \ woven}}$ and a normalized plastic damage $P_{\mathrm{norm.}}$ is defined as $||H|^{\overline{\mathrm{5\%}}}_{\mathrm{after \ unloading}} - |H|^{\overline{\mathrm{5\%}}}_{\mathrm{initial}}  | / |H|^{\overline{\mathrm{5\%}}}_{\mathrm{initial}}$ (the corresponding force--displacement curves are given in Supplementary Fig. S7). The normalized stiffness $K_{\mathrm{0, \ norm.}}$ of the continuous systems is typically below $2$ (average of $1.436$) indicating that weaving only results in a small penalty in stiffness compared to the theoretical maximum of $8$. The woven systems also experience minimal plastic damage with $P_{\mathrm{norm.}}$ typically below $1$, versus continuous systems where $P_{\mathrm{norm.}}$ is typically above $4$. Although Bi5, Bi6, and corners of higher orders have a sign change in global Gaussian curvature compared to Bi3 corners, they similarly experience stretching of ribbons under small axial compression and local buckling of ribbons under large deformation (see Supplementary Fig. S8 and Section S5D for details). Because the local mechanical behaviors are the same, Bi5 and Bi6 corners offer a similar high stiffness and enhanced resilience.

The stiffness and resilience properties are also qualitatively robust with respect to the type of material and the density of ribbons used for woven structures. When woven corners made of wood and stainless steel are tested, they show a slightly higher plastic damage than woven corners made of polymer (Mylar\textsuperscript{\textregistered}) (Fig. \ref{fig:Mechanics}f, bottom panel), yet the normalized plastic damage remains substantially lower compared with their continuous counterparts (see Supplementary Fig. S9 and Section S5E for details). When we increase the ribbon density by reducing the ribbon width in weaving, the woven corners gain better resilience at the cost of a slight decrease in stiffness (see Supplementary Fig. S10 and Section S5F for details). Furthermore, we use different numbers of additional local pins as a proxy to simulate different levels of adhesion and friction of the material. As the number of pins increases, the local buckling of the ribbons becomes increasingly constrained, and we observe a transition from a woven-type behavior to a continuum-type behavior (see Supplementary Fig. S11 and Section S5G for details).

% \begin{figure}
\begin{figure}[!htb] % Use this if the figure floating is not ideal
\centering
\includegraphics[width=17.8cm]{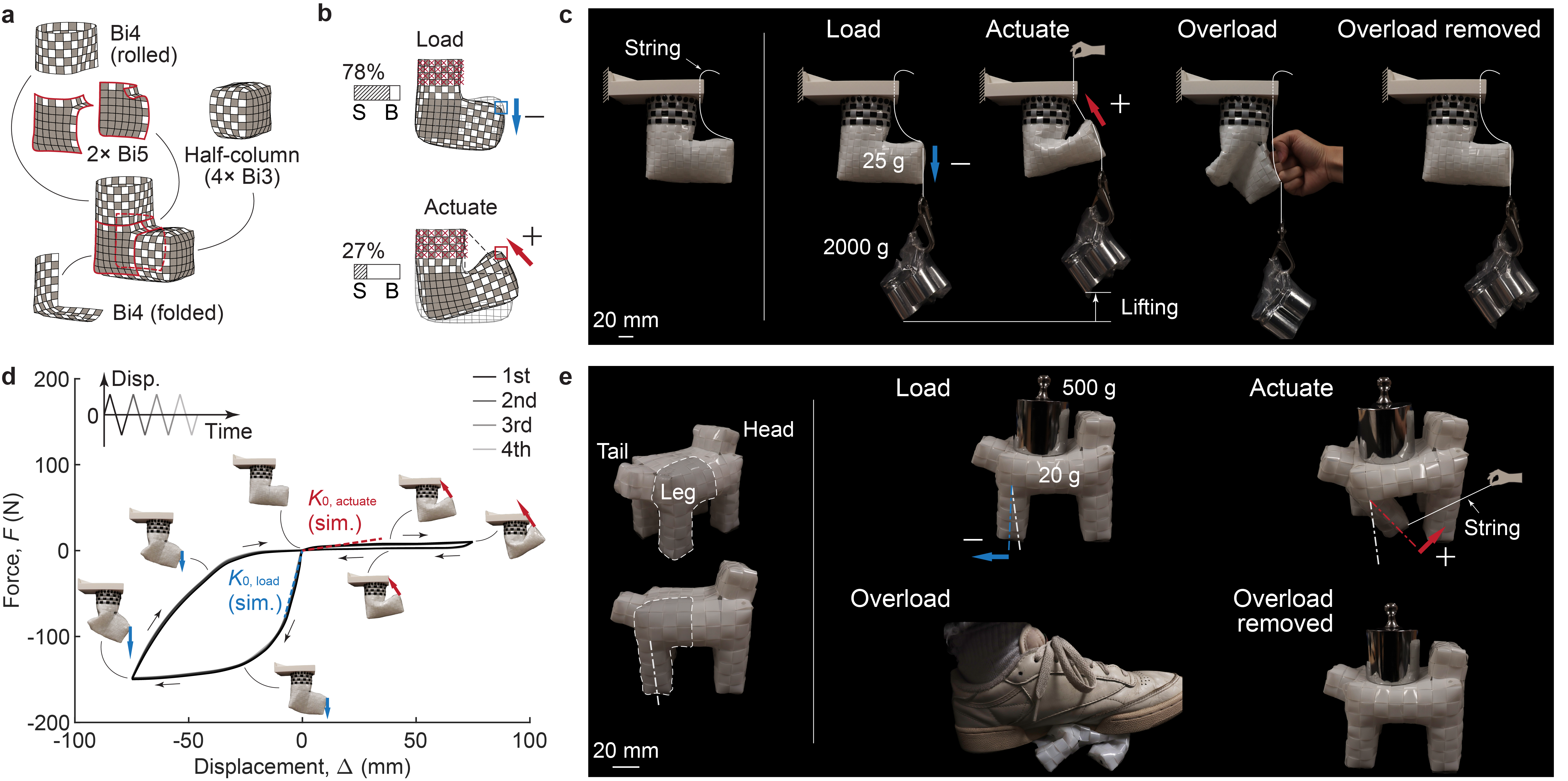}
\caption{\textbf{Damage-resistant woven robotic systems}. \textbf{a}, Woven corner components of a robotic arm. \textbf{b}, Simulated linearly deformed arms under loading and actuation with the corresponding stretching `S' to bending `B' energy ratios. The checkerboard-colored surfaces are the deformed configurations where the deformation is scaled up ($\times $100); the gray frames are initial undeformed shapes; the red crosses indicate boundary nodes where the $x$, $y$, $z$ coordinates are fixed; the two arrows indicate the directions of the load and the colored squares indicate the positions where the load is applied; a unit load is used in both simulations. \textbf{c}, Photographs of the woven arm at the rest state (left) and when we load, actuate, overload it, and remove the overload (right). \textbf{d}, Cyclic force--displacement curves of the arm during four loading and actuation cycles with the simulated loading and actuating stiffnesses juxtaposed (see Supplementary Section S6A for the testing details). The results of the four cyclic tests directly overlap showing no degradation in behavior. \textbf{e}, Photographs of a four-legged woven dog at the rest state (left) and when we load, actuate, overload it, and remove the overload (right).}\label{fig:Robot}
\end{figure}

\begin{figure}[!htb] % Use this if the figure floating is not ideal
\centering
\includegraphics[width=17.8cm]{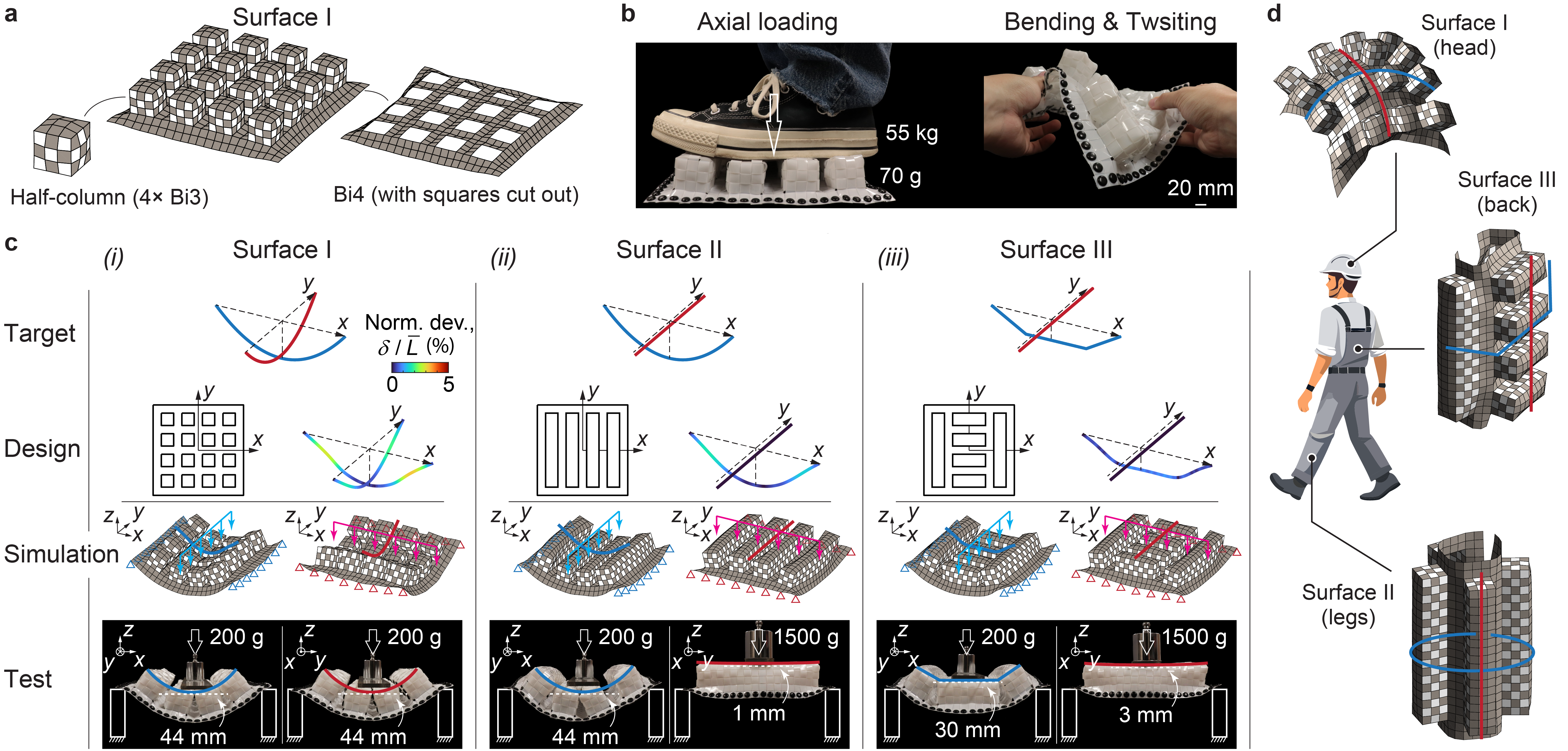}
\caption{\textbf{Woven metasurfaces with tailorable stiffness}. \textbf{a}, Woven corner components of Surface I. \textbf{b}, Photographs of Surface I under axial loading (left) and bending and twisting (right). \textbf{c}, Computational design of three metasurfaces. The first row shows the target bending modes; the second row shows the selected metasurfaces and the normalized deviations between the simulated and target modes $\delta/\overline{L}$ where $\delta$ is the deviation and $\overline{L}$ is the average length of the bending mode curves; the third row shows the simulated linearly deformed shapes of the metasurfaces (scaled $\times 100$) with the colored triangles indicating the nodes fixed in the $z$ direction and the colored arrows indicating the uniformly distributed centerline loads (a unit load is used in all simulations); the fourth row shows a physical test of each fabricated metasurface when simply supported along the two sides and loaded along the centerline; the bending mode is sketched for each case and the displacement shown is measured at the midpoint on top of the structure. \textbf{d}, Demonstration of woven metasurfaces applied as components of an exoskeleton suit.}\label{fig:Metasurface}
\end{figure}

\section*{Functional Structures Built From Woven Corners}
Next, we use the woven corner units presented in Fig. \ref{fig:Intro} as a basis to design functional structures for real-world applications. We use the rigidity provided by axial stiffness to create components that can carry load, and we use the flexibility of bending modes to allow for functional motions or to allow the structure to deform in a desired pattern. All structures created here share the characteristics we observed earlier, in that they are highly resilient to damage even when overloaded and compressed into the densification regime. 

\subsection*{Damage-Resistant Woven Robotic Systems} Woven elbow links such as the one in Fig. \ref{fig:Intro}d(\textit{iv}) can serve as modular robotic arms or legs. We remove the cap on one end of the elbow link and rigidly connect it to a solid cylinder support on top to create a woven robotic arm (Fig. \ref{fig:Robot}a and c). When we vertically load the end of the arm with an object that is 80 times heavier (2 kg versus a 25 g self-weight), the arm remains stiff and minimally deforms. However, when we actuate the arm by pulling a string attached to the end upward and diagonally, the arm easily deforms, making it possible to lift heavy objects efficiently (Fig. \ref{fig:Robot}c). This behavior mimics that of a human arm---it bends easily in one direction but not in the opposite direction. Numerically and experimentally, we show that the stiffness of the arm is drastically different if engaged with a vertical load versus with actuation, as $K_{\mathrm{0, \ load}} / K_{\mathrm{0, \ actuate}} = 25$ (Fig. \ref{fig:Robot}d). This difference occurs because the ribbons are primarily engaged in in-plane stretching when the arm is loaded while they primarily bend and deform out-of-plane when the arm is actuated (Fig. \ref{fig:Robot}b). If the arm accidentally hits a rigid object where it is overloaded and exhibits excessive deformation, it recovers to its initial geometry and can again accomplish the function of lifting a 2 kg weight without any sign of damage (Fig. \ref{fig:Robot}c). Being actuated, overloaded, and actuated again is highly repeatable, as shown in the cyclic force--displacement curves in Fig.~\ref{fig:Robot}d. The repeatability demonstrates the remarkable damage resistance of woven robotic systems and safety of human-woven robot interaction \cite{martinez2014soft,pei2019highly,wang2018improving,pei2019highly}. We next build a four-legged woven dog where each leg is similar to the woven arm module (Fig.~\ref{fig:Robot}e). Similar to the single arm case, the high stiffness for axial loading enables good load bearing capacity, the low bending stiffness enables easy actuation for potential locomotion, and the local elastic buckling of ribbons enables resistance to damage if the structure is overloaded. This concept can be generalized to design other damage-resistant functional robotic systems such as a three-legged crawler as we show in Supplementary Section S6B.

\subsection*{Woven Metasurfaces With Tailorable Stiffness} Following similar length compatibility conditions as those shown in Fig. \ref{fig:Intro}d, we remove sixteen squares from a Bi4 woven sheet and instead connect those regions with half-column ($4\times$ Bi3) modules to form a woven metasurface (which we call Surface I, Fig. \ref{fig:Metasurface}a). The resulting structure supports an axial weight of 55 kg (over 700 times its self-weight) and yet remains flexible in bending and twisting (Fig. \ref{fig:Metasurface}b). The high axial stiffness is achieved because of the shape and orientation of the local Bi3 half-column modules, while the out-of-plane bending flexibility is enabled by global bending of the Bi4 sheet substrate. We characterize the out-of-plane bending modes of the metasurface about the $x$ and $y$ axes by restraining the two edges and loading vertically along the centerline. Surface I is flexible about both the $x$ and $y$ axes (Fig. \ref{fig:Metasurface}c(\textit{i})), but the bending modes of the metasurface can be tailored by modifying the two-dimensional geometric design pattern. Each small square unit on the Bi4 woven sheet can be chosen to be `0' (substrate) or `1' (half-column), so the design space for two-dimensional metasurface patterns is as large as $2^{N\times M}$ where $N\times M$ is the dimension of any rectangular Bi4 flat sheet (see Supplementary Fig. S13\textit{A}). However, to reduce the cost of computational search in the design space, we only modify the pattern by connecting the sixteen individual squares with straight lines along the $x$ and $y$ axes, where we do not allow the lines to intersect or the squares to be isolated (see Supplementary Fig. S13\textit{B}). Within this design space, we have 99 different metasurfaces composed of Bi3 half-columns with rectangular cross-sections. We then identify metasurfaces with bending modes that best match a target shape and stiffness by calculating the degree of similarity between the target and simulated deformation modes (details on the computational approach are given in Supplementary Section S6C). Surface II and Surface III in Fig. \ref{fig:Metasurface}c(\textit{ii}) and (\textit{iii}) are two sample designs based on this method: Surface II is flexible in bending about the $y$ axis, but it is stiff about the $x$ axis (almost rigid in comparison); Surface III provides comparable high stiffness about the $x$ axis, while it is relatively flexible about the $y$ axis and has a piecewise bending mode with a flat region in the middle. These woven metasurfaces can serve in various applications where deformation modes and stiffness need to be tailored. As an example we demonstrate a concept for an exoskeleton suit which adapts to different parts of a human body and provides stiffness as needed (Fig. \ref{fig:Metasurface}d) \cite{xiong2021functional,shin2024woven,yilmaz2024design}. Furthermore, the high axial stiffness of the extruded portions and the excellent resilience of the woven metasurfaces make them well suited for reusable shock absorption, such as preventing injury from external impact as in our example of the human exoskeleton \cite{dong2024suppress,wang2017braided,abtew2019engineering}. The half-column modules can also be assembled from other corner units such as Bi2, and the resulting woven metasurfaces are expected to also have tailorable deformation modes and stiffnesses (see Supplementary Fig. S13\textit{D} and Section S6C). Here, we demonstrate the concept using Bi3 half-column modules to achieve a flat top surface that can support loads.

\section*{Conclusion}
We reveal how the corner topologies in woven baskets and their hierarchical ribbon networks enable a remarkable combination of two conflicting structural properties: high resilience and high stiffness. We show that under large deformations the discontinuity in the woven hierarchy allows for local elastic buckling of the constituent ribbons, which deconcentrates stresses and prevents plastic damage. Therefore, woven corners and assembled woven metamaterials can undergo extreme cyclic deformations and still recover their initial shapes without permanent deformation. Meanwhile, we find that under axial service loads the ribbons are mainly engaged in stretching instead of bending, which makes the stiffness of woven metamaterials close to that of their continuous counterparts. The stiffness ratios between woven metamaterials and comparable continuous structures are consistently closer to `1' than to `1/8'. Here, `1' is the best-case theoretical limit where all ribbons are being loaded only axially, while `1/8' is the worst-case theoretical limit where all ribbons are being bent. These results on stiffness and resilience prove to be robust across different materials, woven-basket geometries, and ribbon width length-scales. 

We generalize the natural plain-woven basket corner Bi3 to a family of corners Bi2--Bi6, where each has a distinct topology yet all share similar mechanical advantages. We present a design approach where these individual corner modules can connect based on dimensional compatibility conditions to assemble into more complex woven metamaterials with conical and hyperbolic geometries. We construct and test woven robotic systems that can be actuated, carry loads, and are damage-resistant to large unintended deformations. We also demonstrate a computational framework for the design of resilient woven metasurfaces where deformation modes can be tailored to be stiff or flexible. We anticipate that the underlying mechanisms that provide both stiffness and resilience to plain-woven baskets can offer a significant opportunity to re-envision the engineering design of lightweight, durable systems across scales and disciplines. Our work also paves the way for future studies where 3D basket weaving is combined with active and electronic materials to enable on-demand spooled fabrication of smart 3D structures and metamaterials with multi-physical functionalities.

% \FloatBarrier % to force the floating stuff (e.g. Figures) to go before this boundary (by Wayne, Aug 12, 2024) 

\section*{Acknowledgements}
The authors acknowledge support from the Air Force Office of Scientific Research under award number FA9550-22-1-0321. The authors also acknowledge helpful discussions with Dr. Yi Zhu. The paper reflects the views and opinions of the authors, and not necessarily those of the funding entities. 

\bibliography{arXiv_Manuscript_June22_25}

\begin{thebibliography}{10}
\urlstyle{rm}
\expandafter\ifx\csname url\endcsname\relax
  \def\url#1{\texttt{#1}}\fi
\expandafter\ifx\csname urlprefix\endcsname\relax\def\urlprefix{URL }\fi
\expandafter\ifx\csname doiprefix\endcsname\relax\def\doiprefix{DOI: }\fi
\providecommand{\bibinfo}[2]{#2}
\providecommand{\eprint}[2][]{\url{#2}}

\bibitem{lan2023woven}
\bibinfo{author}{Lan, L.} \emph{et~al.}
\newblock \bibinfo{journal}{\bibinfo{title}{Woven organic crystals}}.
\newblock {\emph{\JournalTitle{Nature Communications}}} \textbf{\bibinfo{volume}{14}}, \bibinfo{pages}{7582} (\bibinfo{year}{2023}).

\bibitem{wang2017smart}
\bibinfo{author}{Wang, S.} \emph{et~al.}
\newblock \bibinfo{journal}{\bibinfo{title}{Smart wearable kevlar-based safeguarding electronic textile with excellent sensing performance}}.
\newblock {\emph{\JournalTitle{Soft Matter}}} \textbf{\bibinfo{volume}{13}}, \bibinfo{pages}{2483--2491} (\bibinfo{year}{2017}).

\bibitem{parsons2010impact}
\bibinfo{author}{Parsons, E.~M.}, \bibinfo{author}{Weerasooriya, T.}, \bibinfo{author}{Sarva, S.} \& \bibinfo{author}{Socrate, S.}
\newblock \bibinfo{journal}{\bibinfo{title}{Impact of woven fabric: Experiments and mesostructure-based continuum-level simulations}}.
\newblock {\emph{\JournalTitle{Journal of the Mechanics and Physics of Solids}}} \textbf{\bibinfo{volume}{58}}, \bibinfo{pages}{1995--2021} (\bibinfo{year}{2010}).

\bibitem{parsons2013modeling}
\bibinfo{author}{Parsons, E.~M.}, \bibinfo{author}{King, M.~J.} \& \bibinfo{author}{Socrate, S.}
\newblock \bibinfo{journal}{\bibinfo{title}{Modeling yarn slip in woven fabric at the continuum level: Simulations of ballistic impact}}.
\newblock {\emph{\JournalTitle{Journal of the Mechanics and Physics of Solids}}} \textbf{\bibinfo{volume}{61}}, \bibinfo{pages}{265--292} (\bibinfo{year}{2013}).

\bibitem{august2020self}
\bibinfo{author}{August, D.~P.} \emph{et~al.}
\newblock \bibinfo{journal}{\bibinfo{title}{Self-assembly of a layered two-dimensional molecularly woven fabric}}.
\newblock {\emph{\JournalTitle{Nature}}} \textbf{\bibinfo{volume}{588}}, \bibinfo{pages}{429--435} (\bibinfo{year}{2020}).

\bibitem{must2019variable}
\bibinfo{author}{Must, I.}, \bibinfo{author}{Sinibaldi, E.} \& \bibinfo{author}{Mazzolai, B.}
\newblock \bibinfo{journal}{\bibinfo{title}{A variable-stiffness tendril-like soft robot based on reversible osmotic actuation}}.
\newblock {\emph{\JournalTitle{Nature Communications}}} \textbf{\bibinfo{volume}{10}}, \bibinfo{pages}{344} (\bibinfo{year}{2019}).

\bibitem{maziz2017knitting}
\bibinfo{author}{Maziz, A.} \emph{et~al.}
\newblock \bibinfo{journal}{\bibinfo{title}{Knitting and weaving artificial muscles}}.
\newblock {\emph{\JournalTitle{Science Advances}}} \textbf{\bibinfo{volume}{3}}, \bibinfo{pages}{e1600327} (\bibinfo{year}{2017}).

\bibitem{zhang2017rugged}
\bibinfo{author}{Zhang, L.}, \bibinfo{author}{Fairbanks, M.} \& \bibinfo{author}{Andrew, T.~L.}
\newblock \bibinfo{journal}{\bibinfo{title}{Rugged textile electrodes for wearable devices obtained by vapor coating off-the-shelf, plain-woven fabrics}}.
\newblock {\emph{\JournalTitle{Advanced Functional Materials}}} \textbf{\bibinfo{volume}{27}}, \bibinfo{pages}{1700415} (\bibinfo{year}{2017}).

\bibitem{zhao2021soft}
\bibinfo{author}{Zhao, X.} \emph{et~al.}
\newblock \bibinfo{journal}{\bibinfo{title}{Soft fibers with magnetoelasticity for wearable electronics}}.
\newblock {\emph{\JournalTitle{Nature Communications}}} \textbf{\bibinfo{volume}{12}}, \bibinfo{pages}{6755} (\bibinfo{year}{2021}).

\bibitem{backer1951relationship}
\bibinfo{author}{Backer, S.}
\newblock \bibinfo{journal}{\bibinfo{title}{The relationship between the structural geometry of a textile fabric and its physical properties: Part iv: interstice geometry and air permeability}}.
\newblock {\emph{\JournalTitle{Textile Research Journal}}} \textbf{\bibinfo{volume}{21}}, \bibinfo{pages}{703--714} (\bibinfo{year}{1951}).

\bibitem{granger1982weaving}
\bibinfo{author}{Granger-Taylor, H.}
\newblock \bibinfo{journal}{\bibinfo{title}{Weaving clothes to shape in the ancient world: the tunic and toga of the arringatore}}.
\newblock {\emph{\JournalTitle{Textile History}}} \textbf{\bibinfo{volume}{13}}, \bibinfo{pages}{3--25} (\bibinfo{year}{1982}).

\bibitem{wu2019silk}
\bibinfo{author}{Wu, R.} \emph{et~al.}
\newblock \bibinfo{journal}{\bibinfo{title}{Silk composite electronic textile sensor for high space precision 2{D} combo temperature--pressure sensing}}.
\newblock {\emph{\JournalTitle{Small}}} \textbf{\bibinfo{volume}{15}}, \bibinfo{pages}{1901558} (\bibinfo{year}{2019}).

\bibitem{zhang2022elastic}
\bibinfo{author}{Zhang, Y.} \emph{et~al.}
\newblock \bibinfo{journal}{\bibinfo{title}{Elastic fibers/fabrics for wearables and bioelectronics}}.
\newblock {\emph{\JournalTitle{Advanced Science}}} \textbf{\bibinfo{volume}{9}}, \bibinfo{pages}{2203808} (\bibinfo{year}{2022}).

\bibitem{shi2021large}
\bibinfo{author}{Shi, X.} \emph{et~al.}
\newblock \bibinfo{journal}{\bibinfo{title}{Large-area display textiles integrated with functional systems}}.
\newblock {\emph{\JournalTitle{Nature}}} \textbf{\bibinfo{volume}{591}}, \bibinfo{pages}{240--245} (\bibinfo{year}{2021}).

\bibitem{huang2020fiber}
\bibinfo{author}{Huang, L.} \emph{et~al.}
\newblock \bibinfo{journal}{\bibinfo{title}{Fiber-based energy conversion devices for human-body energy harvesting}}.
\newblock {\emph{\JournalTitle{Advanced Materials}}} \textbf{\bibinfo{volume}{32}}, \bibinfo{pages}{1902034} (\bibinfo{year}{2020}).

\bibitem{chai2016tailorable}
\bibinfo{author}{Chai, Z.} \emph{et~al.}
\newblock \bibinfo{journal}{\bibinfo{title}{Tailorable and wearable textile devices for solar energy harvesting and simultaneous storage}}.
\newblock {\emph{\JournalTitle{ACS Nano}}} \textbf{\bibinfo{volume}{10}}, \bibinfo{pages}{9201--9207} (\bibinfo{year}{2016}).

\bibitem{chen2016micro}
\bibinfo{author}{Chen, J.} \emph{et~al.}
\newblock \bibinfo{journal}{\bibinfo{title}{Micro-cable structured textile for simultaneously harvesting solar and mechanical energy}}.
\newblock {\emph{\JournalTitle{Nature Energy}}} \textbf{\bibinfo{volume}{1}}, \bibinfo{pages}{1--8} (\bibinfo{year}{2016}).

\bibitem{xiong2021functional}
\bibinfo{author}{Xiong, J.}, \bibinfo{author}{Chen, J.} \& \bibinfo{author}{Lee, P.~S.}
\newblock \bibinfo{journal}{\bibinfo{title}{Functional fibers and fabrics for soft robotics, wearables, and human--robot interface}}.
\newblock {\emph{\JournalTitle{Advanced Materials}}} \textbf{\bibinfo{volume}{33}}, \bibinfo{pages}{2002640} (\bibinfo{year}{2021}).

\bibitem{buckner2024untethered}
\bibinfo{author}{Buckner, T.~L.}, \bibinfo{author}{Huang, X.} \& \bibinfo{author}{Kramer-Bottiglio, R.}
\newblock \bibinfo{journal}{\bibinfo{title}{Untethered, dynamic robotic fabrics enabled by actively-rigid variable stiffness fibers}}.
\newblock {\emph{\JournalTitle{Advanced Functional Materials}}} \bibinfo{pages}{2404431} (\bibinfo{year}{2024}).

\bibitem{buckner2020roboticizing}
\bibinfo{author}{Buckner, T.~L.}, \bibinfo{author}{Bilodeau, R.~A.}, \bibinfo{author}{Kim, S.~Y.} \& \bibinfo{author}{Kramer-Bottiglio, R.}
\newblock \bibinfo{journal}{\bibinfo{title}{Roboticizing fabric by integrating functional fibers}}.
\newblock {\emph{\JournalTitle{Proceedings of the National Academy of Sciences}}} \textbf{\bibinfo{volume}{117}}, \bibinfo{pages}{25360--25369} (\bibinfo{year}{2020}).

\bibitem{hu2019buckled}
\bibinfo{author}{Hu, X.}, \bibinfo{author}{Dou, Y.}, \bibinfo{author}{Li, J.} \& \bibinfo{author}{Liu, Z.}
\newblock \bibinfo{journal}{\bibinfo{title}{Buckled structures: fabrication and applications in wearable electronics}}.
\newblock {\emph{\JournalTitle{Small}}} \textbf{\bibinfo{volume}{15}}, \bibinfo{pages}{1804805} (\bibinfo{year}{2019}).

\bibitem{suzuki2016wrinkles}
\bibinfo{author}{Suzuki, K.} \& \bibinfo{author}{Ohzono, T.}
\newblock \bibinfo{journal}{\bibinfo{title}{Wrinkles on a textile-embedded elastomer surface with highly variable friction}}.
\newblock {\emph{\JournalTitle{Soft Matter}}} \textbf{\bibinfo{volume}{12}}, \bibinfo{pages}{6176--6183} (\bibinfo{year}{2016}).

\bibitem{rajesh2017experimental}
\bibinfo{author}{Rajesh, M.} \& \bibinfo{author}{Pitchaimani, J.}
\newblock \bibinfo{journal}{\bibinfo{title}{Experimental investigation on buckling and free vibration behavior of woven natural fiber fabric composite under axial compression}}.
\newblock {\emph{\JournalTitle{Composite Structures}}} \textbf{\bibinfo{volume}{163}}, \bibinfo{pages}{302--311} (\bibinfo{year}{2017}).

\bibitem{martinez2023earliest}
\bibinfo{author}{Mart{\'\i}nez-Sevilla, F.} \emph{et~al.}
\newblock \bibinfo{journal}{\bibinfo{title}{The earliest basketry in southern {E}urope: Hunter-gatherer and farmer plant-based technology in {C}ueva de los {M}urci{\'e}lagos ({A}lbu{\~n}ol)}}.
\newblock {\emph{\JournalTitle{Science Advances}}} \textbf{\bibinfo{volume}{9}}, \bibinfo{pages}{eadi3055} (\bibinfo{year}{2023}).

\bibitem{poincloux2023indentation}
\bibinfo{author}{Poincloux, S.}, \bibinfo{author}{Vallat, C.}, \bibinfo{author}{Chen, T.}, \bibinfo{author}{Sano, T.~G.} \& \bibinfo{author}{Reis, P.~M.}
\newblock \bibinfo{journal}{\bibinfo{title}{Indentation and stability of woven domes}}.
\newblock {\emph{\JournalTitle{Extreme Mechanics Letters}}} \textbf{\bibinfo{volume}{59}}, \bibinfo{pages}{101968} (\bibinfo{year}{2023}).

\bibitem{song2023buckling}
\bibinfo{author}{Song, G.-K.} \& \bibinfo{author}{Sun, B.-H.}
\newblock \bibinfo{journal}{\bibinfo{title}{Buckling of ellipsoid grid-shells made of smooth triaxial weaving with naturally in-plane curved ribbons}}.
\newblock {\emph{\JournalTitle{Thin-Walled Structures}}} \textbf{\bibinfo{volume}{191}}, \bibinfo{pages}{111060} (\bibinfo{year}{2023}).

\bibitem{baek2021smooth}
\bibinfo{author}{Baek, C.}, \bibinfo{author}{Martin, A.~G.}, \bibinfo{author}{Poincloux, S.}, \bibinfo{author}{Chen, T.} \& \bibinfo{author}{Reis, P.~M.}
\newblock \bibinfo{journal}{\bibinfo{title}{Smooth triaxial weaving with naturally curved ribbons}}.
\newblock {\emph{\JournalTitle{Physical Review Letters}}} \textbf{\bibinfo{volume}{127}}, \bibinfo{pages}{104301} (\bibinfo{year}{2021}).

\bibitem{ren20213d}
\bibinfo{author}{Ren, Y.} \emph{et~al.}
\newblock \bibinfo{journal}{\bibinfo{title}{3{D} weaving with curved ribbons}}.
\newblock {\emph{\JournalTitle{ACM Transactions on Graphics}}} \textbf{\bibinfo{volume}{40}}, \bibinfo{pages}{127} (\bibinfo{year}{2021}).

\bibitem{lewandowska2017triaxial}
\bibinfo{author}{Lewandowska, U.} \emph{et~al.}
\newblock \bibinfo{journal}{\bibinfo{title}{A triaxial supramolecular weave}}.
\newblock {\emph{\JournalTitle{Nature Chemistry}}} \textbf{\bibinfo{volume}{9}}, \bibinfo{pages}{1068--1072} (\bibinfo{year}{2017}).

\bibitem{peng2024switchable}
\bibinfo{author}{Peng, Y.} \emph{et~al.}
\newblock \bibinfo{journal}{\bibinfo{title}{Switchable pseudo-triaxial structure enabled mechanosensory textiles with ultra-wide detection range for flexible e-wearables}}.
\newblock {\emph{\JournalTitle{Advanced Functional Materials}}} \bibinfo{pages}{2411177} (\bibinfo{year}{2024}).

\bibitem{tu2024origami}
\bibinfo{author}{Tu, G.~W.} \& \bibinfo{author}{Filipov, E.~T.}
\newblock \bibinfo{journal}{\bibinfo{title}{Origami of multi-layered spaced sheets}}.
\newblock {\emph{\JournalTitle{Journal of the Mechanics and Physics of Solids}}} \bibinfo{pages}{105730} (\bibinfo{year}{2024}).

\bibitem{filipov2017bar}
\bibinfo{author}{Filipov, E.}, \bibinfo{author}{Liu, K.}, \bibinfo{author}{Tachi, T.}, \bibinfo{author}{Schenk, M.} \& \bibinfo{author}{Paulino, G.~H.}
\newblock \bibinfo{journal}{\bibinfo{title}{Bar and hinge models for scalable analysis of origami}}.
\newblock {\emph{\JournalTitle{International Journal of Solids and Structures}}} \textbf{\bibinfo{volume}{124}}, \bibinfo{pages}{26--45} (\bibinfo{year}{2017}).

\bibitem{zhu2021rapid}
\bibinfo{author}{Zhu, Y.} \& \bibinfo{author}{Filipov, E.~T.}
\newblock \bibinfo{journal}{\bibinfo{title}{Rapid multi-physics simulation for electro-thermal origami systems}}.
\newblock {\emph{\JournalTitle{International Journal of Mechanical Sciences}}} \textbf{\bibinfo{volume}{202}}, \bibinfo{pages}{106537} (\bibinfo{year}{2021}).

\bibitem{fu2019programmable}
\bibinfo{author}{Fu, K.}, \bibinfo{author}{Zhao, Z.} \& \bibinfo{author}{Jin, L.}
\newblock \bibinfo{journal}{\bibinfo{title}{Programmable granular metamaterials for reusable energy absorption}}.
\newblock {\emph{\JournalTitle{Advanced Functional Materials}}} \textbf{\bibinfo{volume}{29}}, \bibinfo{pages}{1901258} (\bibinfo{year}{2019}).

\bibitem{chen2021reusable}
\bibinfo{author}{Chen, Y.} \& \bibinfo{author}{Jin, L.}
\newblock \bibinfo{journal}{\bibinfo{title}{Reusable energy-absorbing architected materials harnessing snapping-back buckling of wide hyperelastic columns}}.
\newblock {\emph{\JournalTitle{Advanced Functional Materials}}} \textbf{\bibinfo{volume}{31}}, \bibinfo{pages}{2102113} (\bibinfo{year}{2021}).

\bibitem{jiang2023suture}
\bibinfo{author}{Jiang, P.}, \bibinfo{author}{Zhang, S.}, \bibinfo{author}{Yang, H.} \& \bibinfo{author}{Li, Y.}
\newblock \bibinfo{journal}{\bibinfo{title}{Suture interface inspired self-recovery architected structures for reusable energy absorption}}.
\newblock {\emph{\JournalTitle{ACS Applied Materials \& Interfaces}}} \textbf{\bibinfo{volume}{15}}, \bibinfo{pages}{43102--43110} (\bibinfo{year}{2023}).

\bibitem{huttenlocher1993comparing}
\bibinfo{author}{Huttenlocher, D.~P.}, \bibinfo{author}{Klanderman, G.~A.} \& \bibinfo{author}{Rucklidge, W.~J.}
\newblock \bibinfo{journal}{\bibinfo{title}{Comparing images using the {H}ausdorff distance}}.
\newblock {\emph{\JournalTitle{IEEE Transactions on Pattern Analysis and Machine Intelligence}}} \textbf{\bibinfo{volume}{15}}, \bibinfo{pages}{850--863} (\bibinfo{year}{1993}).

\bibitem{Hibbeler2016Mechanics}
\bibinfo{author}{Hibbeler, R.}
\newblock \emph{\bibinfo{title}{Mechanics of Materials, 10th edition}} (\bibinfo{publisher}{Pearson}, \bibinfo{year}{2016}).

\bibitem{peng2024analytic}
\bibinfo{author}{Peng, S.}, \bibinfo{author}{Zhu, Z.} \& \bibinfo{author}{Wei, Y.}
\newblock \bibinfo{journal}{\bibinfo{title}{An analytic solution for bending of multilayered structures with interlayer-slip}}.
\newblock {\emph{\JournalTitle{International Journal of Mechanical Sciences}}} \textbf{\bibinfo{volume}{282}}, \bibinfo{pages}{109642} (\bibinfo{year}{2024}).

\bibitem{martinez2014soft}
\bibinfo{author}{Martinez, R.~V.}, \bibinfo{author}{Glavan, A.~C.}, \bibinfo{author}{Keplinger, C.}, \bibinfo{author}{Oyetibo, A.~I.} \& \bibinfo{author}{Whitesides, G.~M.}
\newblock \bibinfo{journal}{\bibinfo{title}{Soft actuators and robots that are resistant to mechanical damage}}.
\newblock {\emph{\JournalTitle{Advanced Functional Materials}}} \textbf{\bibinfo{volume}{24}}, \bibinfo{pages}{3003--3010} (\bibinfo{year}{2014}).

\bibitem{pei2019highly}
\bibinfo{author}{Pei, Z.}, \bibinfo{author}{Xiong, X.}, \bibinfo{author}{He, J.} \& \bibinfo{author}{Zhang, Y.}
\newblock \bibinfo{journal}{\bibinfo{title}{Highly stretchable and durable conductive knitted fabrics for the skins of soft robots}}.
\newblock {\emph{\JournalTitle{Soft Robotics}}} \textbf{\bibinfo{volume}{6}}, \bibinfo{pages}{687--700} (\bibinfo{year}{2019}).

\bibitem{wang2018improving}
\bibinfo{author}{Wang, Y.}, \bibinfo{author}{Gregory, C.} \& \bibinfo{author}{Minor, M.~A.}
\newblock \bibinfo{journal}{\bibinfo{title}{Improving mechanical properties of molded silicone rubber for soft robotics through fabric compositing}}.
\newblock {\emph{\JournalTitle{Soft Robotics}}} \textbf{\bibinfo{volume}{5}}, \bibinfo{pages}{272--290} (\bibinfo{year}{2018}).

\bibitem{shin2024woven}
\bibinfo{author}{Shin, D.} \emph{et~al.}
\newblock \bibinfo{journal}{\bibinfo{title}{Woven fabric muscle for soft wearable robotic application using two-dimensional zigzag shape memory alloy actuator}}.
\newblock {\emph{\JournalTitle{Soft Robotics}}}  (\bibinfo{year}{2024}).

\bibitem{yilmaz2024design}
\bibinfo{author}{Yilmaz, A.~F.} \emph{et~al.}
\newblock \bibinfo{journal}{\bibinfo{title}{Design and scalable fast fabrication of biaxial fabric pouch motors for soft robotic artificial muscle applications}}.
\newblock {\emph{\JournalTitle{Advanced Intelligent Systems}}} \bibinfo{pages}{2300888} (\bibinfo{year}{2024}).

\bibitem{dong2024suppress}
\bibinfo{author}{Dong, X.}, \bibinfo{author}{Tu, G.}, \bibinfo{author}{Hu, L.} \& \bibinfo{author}{Peng, Z.}
\newblock \bibinfo{journal}{\bibinfo{title}{Suppress chatter in milling of thin-walled parts via fixture with active support}}.
\newblock {\emph{\JournalTitle{Journal of Vibration and Control}}} \textbf{\bibinfo{volume}{30}}, \bibinfo{pages}{1286--1296} (\bibinfo{year}{2024}).

\bibitem{wang2017braided}
\bibinfo{author}{Wang, C.}, \bibinfo{author}{Roy, A.}, \bibinfo{author}{Chen, Z.} \& \bibinfo{author}{Silberschmidt, V.~V.}
\newblock \bibinfo{journal}{\bibinfo{title}{Braided textile composites for sports protection: energy absorption and delamination in impact modelling}}.
\newblock {\emph{\JournalTitle{Materials \& Design}}} \textbf{\bibinfo{volume}{136}}, \bibinfo{pages}{258--269} (\bibinfo{year}{2017}).

\bibitem{abtew2019engineering}
\bibinfo{author}{Abtew, M.~A.}, \bibinfo{author}{Boussu, F.}, \bibinfo{author}{Bruniaux, P.}, \bibinfo{author}{Loghin, C.} \& \bibinfo{author}{Cristian, I.}
\newblock \bibinfo{journal}{\bibinfo{title}{Engineering of 3{D} warp interlock p-aramid fabric structure and its energy absorption capabilities against ballistic impact for body armour applications}}.
\newblock {\emph{\JournalTitle{Composite Structures}}} \textbf{\bibinfo{volume}{225}}, \bibinfo{pages}{111179} (\bibinfo{year}{2019}).

\end{thebibliography}

\end{document}